\documentclass{article}

\usepackage{arxiv}

\usepackage[utf8]{inputenc} 
\usepackage[T1]{fontenc}    
\usepackage{hyperref}       
\usepackage{url}            
\usepackage{booktabs}       
\usepackage{amsfonts}       
\usepackage{nicefrac}       
\usepackage{microtype}      
\usepackage{graphicx}
\usepackage{doi}
\usepackage{multirow}
\usepackage{amssymb}
\usepackage{amsmath}
\usepackage{amsthm}

\newtheorem{remark}{Remark}

\title{A dependent circular-linear model for multivariate biomechanical data: Ilizarov ring fixator study} 

\author{Priyanka Nagar \\
Department of Statistics and Actuarial Science \\
Stellenbosch University \\
Stellenbosch \\
South Africa \\
\texttt{pnagar@sun.ac.za} \\
\AND
Andriette Bekker \\
Department of Statistics \\
University of Pretoria \\
Pretoria \\
South Africa \\
\AND
Mohammad Arashi \\
Department of Statistics \\
Ferdowsi University of Mashhad \\
Mashhad \\
Iran \\
\AND
Cor-Jacques Kat \\
Department of Mechanical and Aeronautical Engineering \\
University of Pretoria \\
Pretoria \\
South Africa \\ 
\AND
Annette-Christi Barnard \\
Walk-A-Mile Centre for Advanced Orthopaedics \\
Centurion \\ 
South Africa \\
}

\renewcommand{\shorttitle}{Nagar et al.}

\begin{document}
\maketitle

\begin{abstract}
Biomechanical and orthopaedic studies frequently encounter complex datasets that encompass both circular and linear variables. In most cases the circular and linear variables are (i) considered in isolation with dependency between variables neglected and (ii) the cyclicity of the circular variables disregarded resulting in erroneous decision making. Given the inherent characteristics of circular variables, it is imperative to adopt methods that integrate directional statistics to achieve precise modelling. This paper is motivated by the modelling of biomechanical data, i.e., the fracture displacements, that is used as a measure in external fixator comparisons. We focus on a data set, based on an Ilizarov ring fixator, comprising of six variables. A modelling framework applicable to the 6D joint distribution of circular-linear data based on vine copulas is proposed. The pair-copula decomposition concept of vine copulas represents the dependence structure as a combination of circular-linear, circular-circular and linear-linear pairs modelled by their respective copulas. This framework allows us to assess the dependencies in the joint distribution as well as account for the cyclicity of the circular variables. Thus, a new approach for accurate modelling of mechanical behaviour for Ilizarov ring fixators and other data of this nature is imparted.
\end{abstract}


\section{Introduction}\label{sec: intro}

External fixators is a medical instrument used to immobilise fractures or heavy damage to the bone structure. Healing of a fracture is influenced by the amount of strain at the fracture site \cite{glatt2017concert}. The strain is related to the amount of motion of the fracture under loading. The motion of the fracture relies on the combination of rings, bars, pins and wires. Different configurations of external fixators can be compared by measuring the displacement of the fracture. Fracture healing is inevitably influenced by the complex interplay of biology and biomechanics i.e. inter-fragmentary motion and inter-fragmentary biomechanics. The stiffness of a construct of an external fixator is determined by the configuration of the hardware. The construct's configuration may lead to different inter-fragmentary motion and it is therefore important to obtain the most appropriate construct for optimal healing. Iobst et al. \cite{iobst2023review} provided an overview of the various circular external fixators based on current choices available within the field, accompanied by a comprehensive comparative assessment of seven prevalent hexapod circular external fixator systems. Pertinent attributes pertaining to the hardware, software components, and educational potentials of each system are meticulously delineated. Noteworthy is the systematic refinement of information procured from system manufacturers, subject to rigorous and unbiased editorial scrutiny to align with the structural requirements of this review, devoid of any extraneous subjective commentary or recommendations. The authors of the review contend that the development of varied systems has had a positive impact on the advancement of this technical domain.

In the literature various studies can be found which compare configurations of constructs to determine strain at a fracture site and thus the most appropriate configuration for optimal healing\cite{fenton2021comparative,  gessmann2011influence, henderson2016biomechanical, corona2022outcomes, watts2023comparative}. To the best of the author's knowledge, all studies in this area follow a similar statistical procedure: the assumption of independence among the variables and all variables are considered linear variables. There are two main shortfalls of this approach: (i) disregard of the dependence structure between the variables which may impair the accuracy of the model and (ii) the cyclicity of the rotational variables is neglected and mistreated as a linear variable which may produce misleading results. In numerous studies, the analysis of variance (ANOVA) test is frequently employed to facilitate comparisons among various fixator frames, as supported by research conducted by authors such as Corona et al. \cite{corona2022outcomes} and Watts et al. \cite{watts2023comparative}, among others. These investigations are grounded in the analysis of clinical data stemming from patient cohorts. In contrast, this study diverges from this approach by utilising a virtual model to generate simulated data, which emulates the behaviour of the fixator frame under scrutiny. 

In the biomechanical domain, studies including the use of directional statistical techniques have been limited to the univariate and bivariate setting\cite{rivest2001directional, rivest2008directional,pataky2020using, telschow2021functional}. The primary objective of the study by Pataky et al. \cite{pataky2020using} was to conduct a comparative analysis between directional analysis and uni- as well as multivariate Cardan analysis with regard to representative joint kinematic data collected during gait. In the work authored by Telschow et al. \cite{telschow2021functional}, they addressed the issue of gait reproducibility by examining rotations of the tibia and femur at the knee joint. This investigation takes into account both the spatial influence of marker placement and the temporal variability introduced by different walking speeds in the experimental setup. 
The study conducted by Rivest et al. \cite{rivest2008directional} centres their attention on estimating the orientations of the two rotation axes at the ankle joint. While Rivest et al. \cite{rivest2001directional} developed a score statistic to evaluate the adequacy of the fixed‐axis model. Furthermore, they proposed techniques to address the auto-correlation of errors among adjacent data points. These studies mentioned above represent the most immediate applications of directional statistics in the biomechanical field, indicating the existence of a significant research gap that needs to be addressed. The biomechanical domain is rich with angular data, however, the use of directional statistics techniques are overlooked.

The remainder of this paper follows as, Section \ref{sec:backgroud} provides some background on the statistical techniques required to build our proposed model defined in Section \ref{sec: methods}. Section \ref{sec: application} contains the results and discussion of the data application. Section \ref{sec: conclusion} concludes with some final remarks.

\section{Background}\label{sec:backgroud} 

There are a large number of applications which require the analysis of data not realised in the Euclidean space, but rather on some manifold (circles, spheres, hyperspheres, cylinders, torus etc.). Directional statistics constitutes a specialised branch within statistical analysis that focuses on angular observations. An inherent challenge in handling directional data lies in the non-linearity of the sample space, typically represented as a circle or circular manifold. This unique feature has garnered increased attention in the past two decades, primarily driven by the increase of data and the corresponding demand for tailored statistical methodologies, as highlighted by Ley et al. \cite{ley2017modern}. Existing directional models have been summarised in the works of Mardia et al. \cite{mardia2009directional}, Ley et al. \cite{ley2017modern}, and Pewsey et al. \cite{pewsey2021recent}. A fundamental query arises when considering the necessity of directional statistics: "Is there a need for specialised techniques due to the curvature of the sample space? Why are standard linear statistical methods insufficient?". Mardia et al. \cite{mardia2009directional} provide a simple example by considering the metric of the mean to illustrate why it is necessary to account for the curvature of the sample space and how assuming linear techniques leads to inaccurate results. However, in many studies, the cyclicity of a circular variable is often neglected and mistreated as a linear variable \cite{wang2021circular, ZHENG201910, SOLARI201648, LEGUEY2019240}. The biomechanical domain is one example where angular data is prevalent, however, the use of directional statistics techniques are overlooked.
 
 Modelling of circular-linear (C-L) data is limited to the bivariate C-L joint distribution. Various bivariate C-L distributions have been proposed and studied, specifically focusing on meteorology and climatology studies. A well-known method to obtain a C-L distribution is via the conditional modelling approach. Mardia et al.\cite{mardia1978model} combined the von Mises and normal distribution to obtain a joint model with conditional independence. Thereafter many other models have been proposed, however, this approach restricts the choice of the marginal distribution and may neglect the dependence structure between the variables. This concern becomes even more complicated when extending the C-L distribution to the multivariate setting. Even with the assumption of independence among the circular and linear variables, extending the C-L distribution past the bivariate setting is difficult due to the normalising constant being intractable in most cases as well as the restricted circumstances for the normalising constant to be approximated. Thus, resulting in additional complexity with regards to efficient estimation methods. An attempt to extend the model proposed by Mardia et al.\cite{mardia1978model} to the multivariate setting is given by Luengo-Sanchez\cite{Luengo-Sanchez2016}. In these studies, dependencies between linear variables were considered however, every circular variable was considered independent of all the other variables.
 
 One of the most commonly used C-L distributions is the angular-linear model proposed by Johnson-Wehlry\cite{johnson1978some}, known as the JW model. This model was constructed using copulas. Copulas are used to link the marginal distributions of variables to their multivariate distribution\cite{nelsen2007introduction}. Using copulas to build multivariate distributions is a flexible and convenient approach. In the linear domain, a substantial amount of literature can be found on multivariate copulas\cite{nelsen2007introduction, joe2014dependence}. However, these copulas cannot directly be applied to our problem as we need to consider the directional variables and account for their cyclicity. In the directional domain at present only bivariate copulas have been studied for circular-circular (C-C) variables\cite{jones2015class, kato2018circulas} and circular-linear (C-L) variables. The JW model being the most extensively considered and used. The JW model allows the distribution of the C-L variable pair to be written as a function of the marginals and the JW copula function. Since copula functions in the directional domain are limited to the bivariate case an alternative model needs to be considered to obtain our joint 6D model.
 
 A possible solution for the shortfall of multivariate C-L copulas is vine copulas. The origin of vine copulas stem from the hierarchical copula-based structure, specifically the pair copula construction\cite{joe1996m} and then investigated by Bedford et al.\cite{bedford2001probability, bedford2002vines}. There has been a growing interest in vine copulas due to their flexibility and intuitive decomposition. Vine copulas allow the construction of multivariate joint probability distributions by decomposing the multivariate function into simple building blocks comprising of bivariate copulas (pair copulas) based on conditional probabilities\cite{aas2009pair}. Hence, the need for a multivariate C-L copula function is avoided. A vine is usually divided into D-vine and canonical vine (C-vine) structures where the latter is more suitable for situations where a variable is key in controlling the dependency. A $d-$dimensional vine contains $(d-1)$ trees, where the trees are represented with nodes and edges. The nodes are used to determine the edge label. The edge label corresponds to the subscript of each pair copula density and each edge corresponds to one pair copula. Thus the final $d$-dimensional copula can be obtained from the product of the pair copula densities given by all the edges.
 
 By considering vine copulas with our linear-linear (L-L) variable pairs, C-C variable pairs and C-L variable pairs we can construct a multivariate joint probability model while also taking into account the cyclicity of the directional variables. For the trivariate case of a circular-linear-linear joint model this approach has proven to be useful in meteorology and oceanography\cite{wang2021circular, heredia2019modeling}. 
 
 With the flexibility offered by vine copulas comes an increase in the complexity in larger dimensions. This results in the computational effort required to estimate all the parameters growing exponentially with dimension. Given that our data set is 6D we consider a truncated vine copula. Truncated vine copulas have been proposed by Kurowicka\cite{kurowicka2010optimal} and Brenchmann et al.\cite{brechmann2012truncated}. Truncated vine copulas are helpful as they can be constructed by using only pair-wise copulas and a lower number of conditional pair-wise copulas. This method is applied by considering a vine copula structure where all pair-wise copulas with conditioning set equal to or larger than $K$ are replaced with independence copulas. Various approaches for obtaining the optimal $K$ exists\cite{kurowicka2010optimal, brechmann2012truncated}.
 
 To summarise: in this paper we propose a modelling framework applicable to the 6D joint distribution.  Our proposed model makes use of directional statistics to account for the cyclicity of the rotational variables and is constructed based on truncated vine copulas. The pair copula decomposition concept of vine copulas represents the dependence structure as a combination of L-L, C-C and C-L pairs modelled by their respective copulas. This allows us to assess the dependencies in the joint distribution. An advantage of using vine copulas is the flexibility to build multivariate distributions via bivariate copulas that model the dependence between pairs of random variables. The truncation of the vine copula assists with the computation effort required to estimate our model. 
 
\section{Data and Motivation}\label{sec: motivation}

\begin{figure*}[ht!]
\centering
\includegraphics[width=\textwidth]{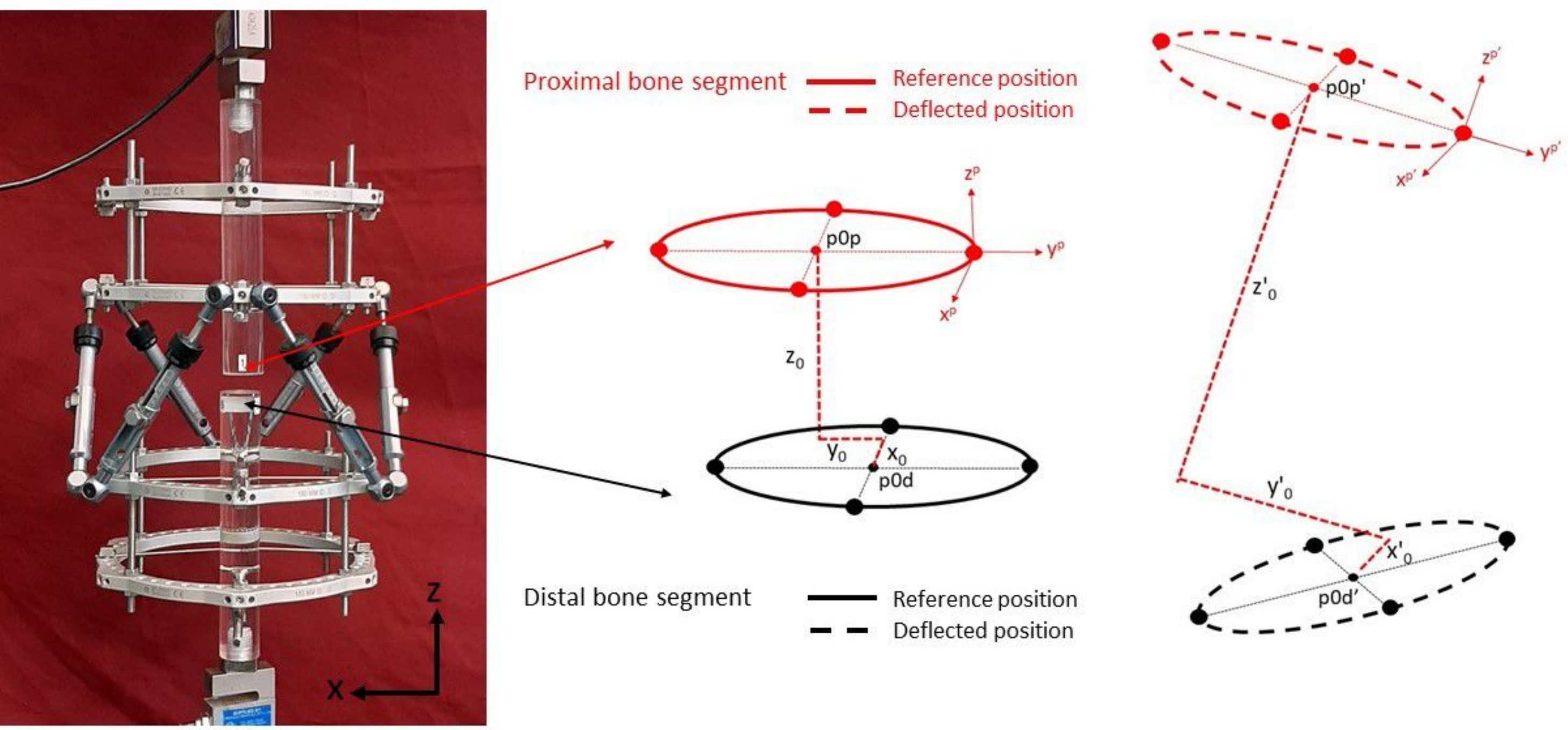}
\caption{\label{fig:virtual model}External fixator on left. Fracture ends at reference position in middle. Fracture ends at deflected position on right.}
\end{figure*}

Let $X, Y$ and $Z$ denote the linear (translational) variables defined on $\mathbb{R}$ where $X$ is the displacement in the $x-axis$, $Y$ the displacement in the $y-axis$ and $Z$ the displacement in the $z-axis$. Let $\Theta_X, \Theta_Y$ and $\Theta_Z$ denote the circular (rotational) variables defined on the unit circle $\mathbb{S}^1$ where $\Theta_X$ is the angular displacement about the $x-axis$, $\Theta_Y$ the angular displacement about the $y-axis$, and $\Theta_Z$ the angular displacement about the $z-axis$. In this study the motion of the fracture is quantified by the displacement of the proximal and distal fracture ends. The translational and rotational motion is measured relative to an axis system fixed to, and rotating with, the proximal bone segment (i.e. $x^p-y^p-z^p$ in Figure \ref{fig:virtual model}). The translational motion is defined by considering the centroid of the two fracture surfaces (i.e. p0p and p0d in Figure \ref{fig:virtual model}). Each motion component is calculated by subtracting the reference position from the deflected position. 

For this application we consider the fracture displacement data comprising of the translational and rotational variables for two different constructs of an Ilizarov ring fixator which we refer to as configuration 1 and configuration 2.
In Table \ref{tab:lin stats}, the descriptive statistics for the translational (linear) variables of configuration 1 and configuration 2 are given; specifically, the mean, standard deviation (sd), interquartile range (IQR), skewness, and kurtosis.

\begin{table*}[ht!]
\small\sf\centering
\caption{\label{tab:lin stats}Descriptive statistics of the translational variables for each configuration.}
\begin{tabular}{lllllll}
\hline
\textbf{Configuration}      & \textbf{Variable} & \textbf{Mean} & \textbf{sd} & \textbf{IQR} & \textbf{skewness} & \textbf{kurtosis} \\
\hline
\multirow{3}{*}{\textbf{1}} & \textbf{$X$}      & -0.2708       & 0.4475      & 0.5192       & -0.3707           & -0.0128           \\ \cline{2-7} 
                            & \textbf{$Y$}      & 0.2948        & 0.4565      & 0.5646       & -0.4791           & 0.2707            \\ \cline{2-7} 
                            & \textbf{$Z$}      & 5.1039        & 3.6230      & 6.9932       & 0.0084            & -1.4289           \\ \hline
\multirow{3}{*}{\textbf{2}} & \textbf{$X$}      & 0.0455        & 0.4760      & 0.5915       & -0.7638           & 2.573             \\ \cline{2-7} 
                            & \textbf{$Y$}      & 1.5551        & 2.2361      & 3.1523       & 1.0108            & -0.0973           \\ \cline{2-7} 
                            & \textbf{$Z$}      & 6.7993        & 4.4999      & 8.9947       & -0.0029           & -1.5106           \\ 
\hline
\end{tabular}
\end{table*}

In Table \ref{tab:circ stats}, the values of the main circular statistics\cite{mardia2009directional} are given for the rotational variables of configuration 1 and configuration 2. Specifically, the mean resultant length ($\bar{r}$), mean direction ($\bar{\theta}$), circular standard deviation ($V_{\Theta}$), circular skewness ($\hat{s}$) and circular kurtosis ($\hat{k}$) are shown. The unit of measurement for the circular variables is considered as radians.

\begin{table*}[ht!]
\small\sf\centering
\caption{\label{tab:circ stats}Descriptive statistics of the rotational variables for each configuration.}
\begin{tabular}{lllllll}
\hline
\textbf{Configuration}      & \textbf{Variable}   & \textbf{$\bar{r}$} & \textbf{$\bar{\theta}$} & \textbf{$V_{\Theta}$} & \textbf{$\hat{s}$} & \textbf{$\hat{k}$} \\
\hline
\multirow{3}{*}{\textbf{1}} & \textbf{$\Theta_X$} & 0.9999             & 0.0011                  & 0.0143                & 0.3286             & -1.1368            \\ \cline{2-7} 
                            & \textbf{$\Theta_Y$} & 0.9999             & -0.0056                 & 0.0127                & -0.1026            & 0.2556             \\ \cline{2-7} 
                            & \textbf{$\Theta_Z$} & 0.9999             & 0.0008                  & 0.0119                & 0.0048             & -1.2235            \\ \hline
\multirow{3}{*}{\textbf{2}} & \textbf{$\Theta_X$} & 0.9992             & -0.0642                 & 0.0397                & 2.7537             & 6.0765             \\ \cline{2-7} 
                            & \textbf{$\Theta_Y$} & 0.9999             & 0.0166                  & 0.0165                & -0.0424            & 0.1532             \\ \cline{2-7} 
                    & \textbf{$\Theta_Z$} & 0.9998             & 0.0194                  & 0.0209                & -1.7600            & 1.4948   \\ 
\hline
\end{tabular}
\end{table*}

\begin{figure*}[ht!]
\centering
\includegraphics[width=\textwidth]{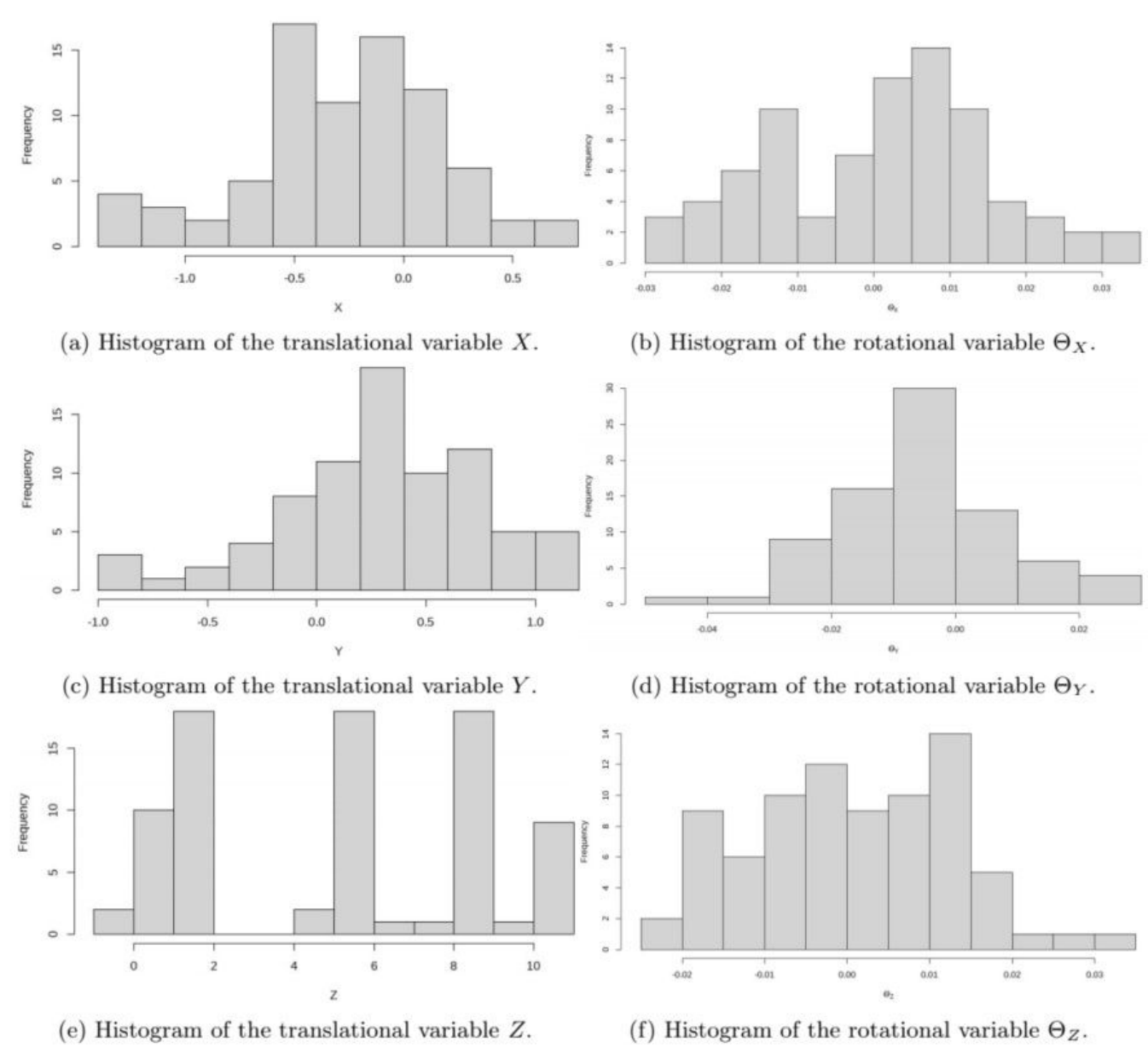}
\caption{\label{fig: c1 hist}Histograms of the data of the translational (left) and rotational (right) variables for configuration 1.}
\end{figure*}

\begin{figure*}[ht!]
\centering
\includegraphics[width=\textwidth]{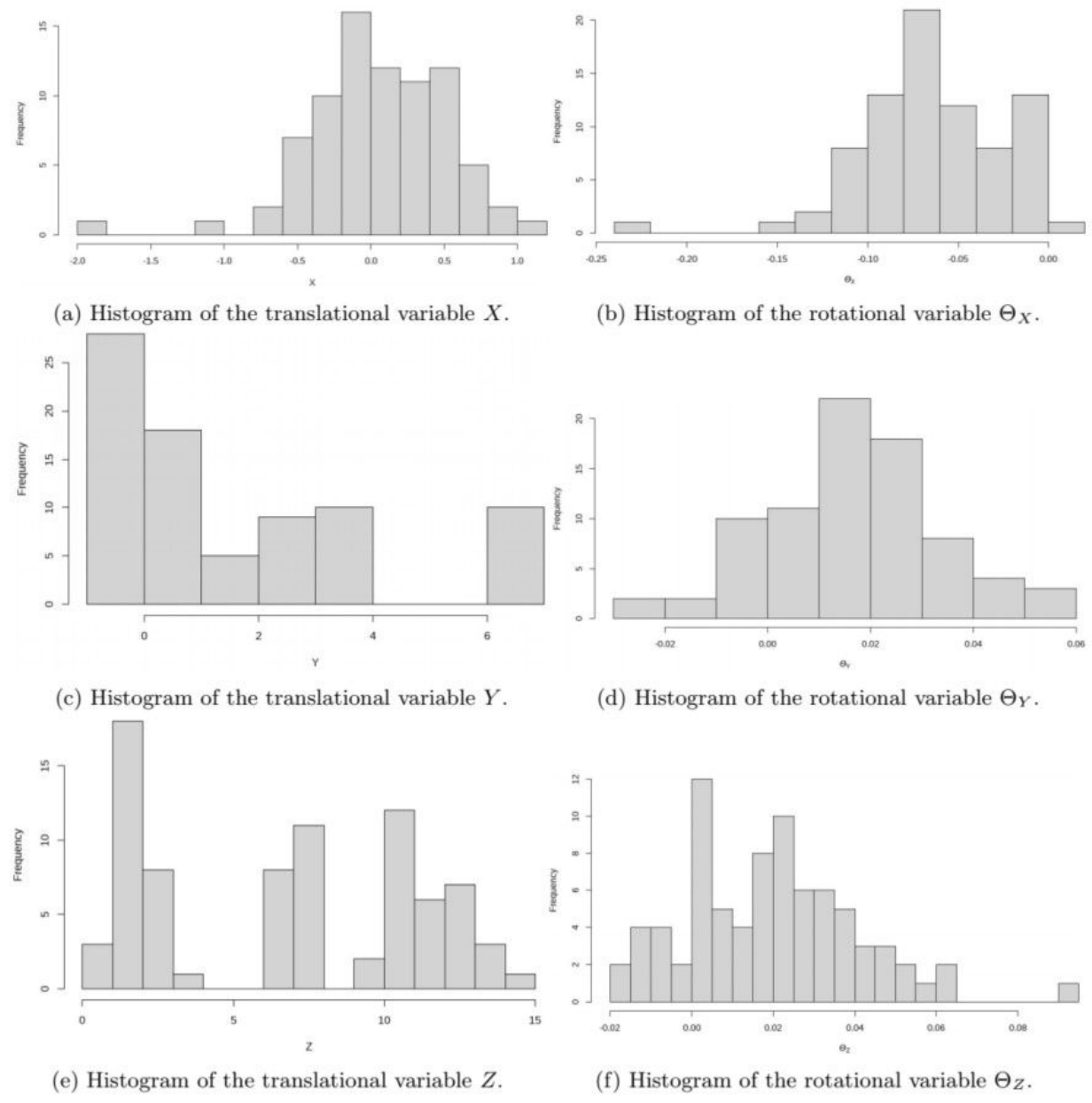}
\caption{\label{fig: c2 hist}Histograms of the data of the translational (left) and rotational (right) variables for configuration 2.}
\end{figure*}

In Figures \ref{fig: c1 hist} and \ref{fig: c2 hist}, the data plots of the translation and rotational variables for configuration 1 and configuration 2 are provided, respectively. From Figures \ref{fig: c1 hist} and \ref{fig: c2 hist}, bimodality and slight skewness are observed in some of the data plots indicating the need for a model that can account for multimodality, for example, the use of finite mixture models for specific variables. Based on these model requirements, a copula approach where the marginal distributions can be specified separately from the dependence structure is a useful technique for building a joint model. From the results in Table {\ref{tab:circ stats}} and the histograms in Figures {\ref{fig: c1 hist}} and {\ref{fig: c2 hist}}, we note that the rotational variables are highly concentrated which raises questions on the assumption of the periodicity of the variables. However, due to the nature of the variables we consider them to be circular in the modelling approach as the concentration might differ for various manufacturers and configurations but the nature of the variable remains unchanged.

\begin{figure*}[ht!]
\centering
\includegraphics[width=\textwidth]{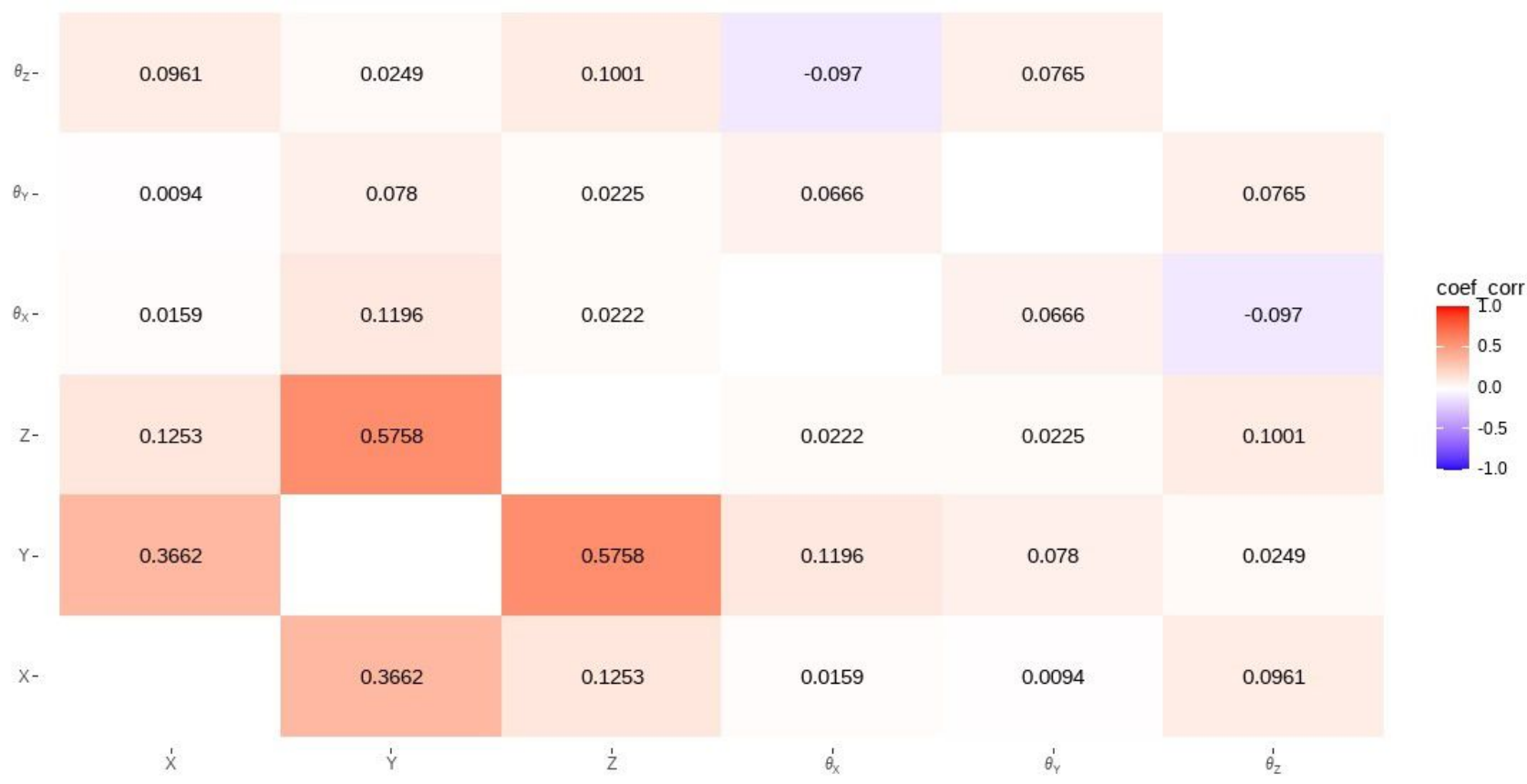}
\caption{\label{fig: corr plot c1}Correlation plot of the paired translational variables, paired rotational variables and translation-rotational variables for configuration 1.}
\end{figure*}

\begin{figure*}[ht!]
\centering
\includegraphics[width=\textwidth]{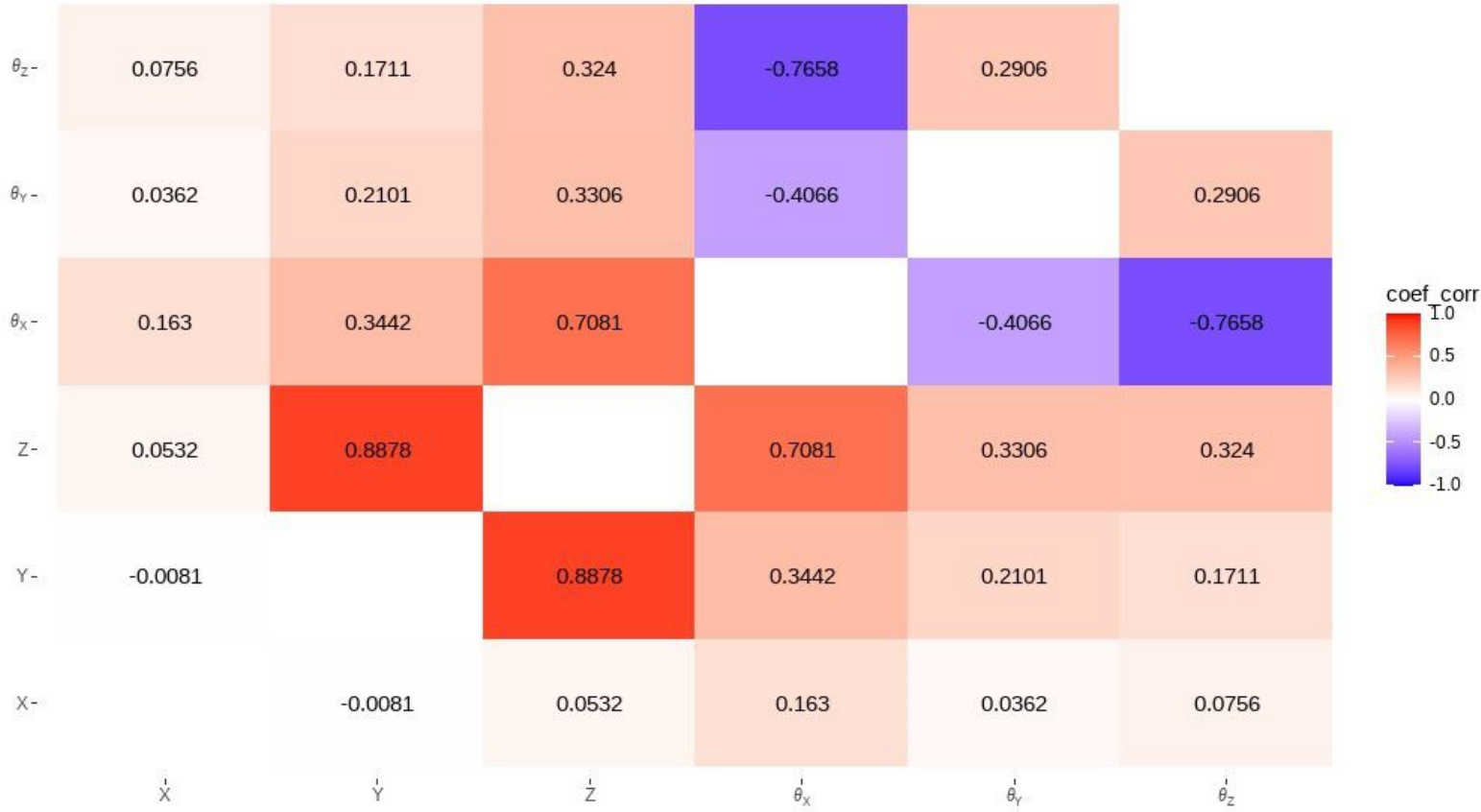}
\caption{\label{fig:corr plot c2}Correlation plot of the paired translational variables, paired rotational variables and translation-rotational variables for configuration 2.}
\end{figure*}

Pearson's correlation coefficient was considered for the paired translational variables. The bivariate relationship between the translational and rotational variables is measured by the linear-circular correlation coefficient\cite{mardia2009directional}, $\rho_{x,\theta}$, as
\begin{equation}
\rho_{x,\theta} = \sqrt{\frac{\rho^{2}_{xc} + \rho^{2}_{xs} - 2\rho_{xc}\rho_{xs}\rho_{cs}}{1 - \rho^{2}_{cs}}},
\end{equation}
where $\rho_{xc} = cor(x,\cos\theta)$, $\rho_{xs} = cor(x,\sin\theta)$, and $\rho_{cs} = cor(\cos\theta,\sin\theta)$ are the sample correlation coefficients. For the paired rotational variables the correlation coefficient\cite{mardia2009directional} is defined as

\begin{equation}
    \resizebox{.9\hsize}{!}{$\rho_{\theta, \phi} = \sqrt{\frac{\left(\rho^2_{cc} + \rho^2_{cs} + \rho^2_{sc} + \rho^2_{ss} \right) + 2\left(\rho_{cc}\rho_{ss} + \rho_{cs}\rho_{sc}\right)\rho_{1}\rho_{2} - 2\left(\rho_{cc}\rho_{cs} + \rho_{sc}\rho_{ss}\right)\rho_{2} -  2\left(\rho_{cc}\rho_{sc} + \rho_{cs}\rho_{ss}\right)\rho_{1}}{\left(1-\rho^2_{1}\right)\left(1- \rho^2_{2}\right)}}$},
\end{equation}
where $\rho_{cc} = cor(\cos\theta, \cos\phi)$, $\rho_{ss} = cor(\sin\theta, \sin\phi)$, $\rho_{cs} = cor(\cos\theta, \sin\phi)$, $\rho_{sc} = cor(\sin\theta, \cos\phi)$, $\rho_{1} = cor(\cos\theta, \sin\theta)$ and $\rho_{2} = cor(\cos\phi, \sin\phi)$ are the sample correlation coefficients.
The correlation plots for configuration 1 and configuration 2 are given in Figure \ref{fig: corr plot c1} and Figure \ref{fig:corr plot c2} respectively. From Figures \ref{fig: corr plot c1} and \ref{fig:corr plot c2} and Tables \ref{tab:corr c1} and \ref{tab:corr c2} it is observed that the assumption of independence is violated by the data. Thus, a joint dependent model is required.
    
A significance test was performed to evaluate the necessity of accounting for a dependence structure between the variables. Table \ref{tab:corr c1} and Table \ref{tab:corr c2} provide the significant (at a 5\% significance level) correlation coefficients for configuration 1 and configuration 2, respectively, where the null hypothesis assumes independency among the respective variables. The results from the correlation tests further emphasise the need for a joint model that accounts for dependencies.

\begin{table}[ht!]
\centering
\caption{\label{tab:corr c1}Paired variables for configuration 1 with correlation coefficients significant at a $5$\% significance level.}

\begin{tabular}{llll}\hline
\multicolumn{2}{l}{\textbf{Paired variables}} & $\mathbf{\rho}$ & \textbf{p-value} \\ \hline
$X$                   & $Y$                  & 0.3662          & 0.0017           \\
$Y$                   &  $Z$                  & 0.5758          & 0.000            \\
$Y$                   & $\Theta_X$                  & 0.1196          & 0.0001           \\
$Y$                   & $\Theta_Y$                  & 0.078           & 0.0024           \\
$X$                   & $\Theta_Z$                  & 0.0961          & 0.0006           \\ 
$Z$                   & $\Theta_Z$                  & 0.1001          & 0.0004          \\\hline
\end{tabular}
\end{table}

\begin{table}[ht!]
\centering
\caption{\label{tab:corr c2}Paired variables for configuration 2 with correlation coefficients significant at a $5$\% significance level.}
\begin{tabular}{llll}\hline
\multicolumn{2}{l}{\textbf{Paired variables}} & $\mathbf{\rho}$ & \textbf{p-value} \\ \hline
$Y$                  & $Z$                   & 0.8878          & 0.000            \\
$\Theta_X$                  & $\Theta_Y$                  & -0.4066         & 0.0218           \\
$\Theta_X$                  & $\Theta_Z$                  & -0.7658         & 0.0007           \\
$X$                   & $\Theta_X$                  & 0.163           & 0.000            \\
$Y$                   & $\Theta_X$                  & 0.3442          & 0.000            \\
$Z$                   & $\Theta_X$                  & 0.7081          & 0.000            \\
$Y$                   & $\Theta_Y$                  & 0.2101          & 0.000            \\
$Z$                   & $\Theta_Y$                  & 0.3306          & 0.000            \\
$X$                   & $\Theta_Z$                  & 0.0756          & 0.0029           \\
$Y$                  & $\Theta_Z$                  & 0.1711          & 0.000            \\
$Z$                   & $\Theta_Z$                  & 0.324           & 0.000           \\ \hline
\end{tabular}
\end{table}

In this paper we propose a modelling framework applicable to the 6D joint distribution to address the above shortcomings. For modelling the displacement of a fracture site as described before, we consider the use of the six variables of which three are translational and three rotational. 

\section{Methodology}\label{sec: methods}

In this section we define the proposed modelling framework applicable to the 6D joint distribution. The linear and circular probability density functions (PDFs) considered are given as well as the copula functions. 

\subsection{Linear and circular distributions}
If $X,Y$ and $Z$ denote the linear (translational) variables defined on $\mathbb{R}$, let the PDFs be denoted as $f_X(x), f_Y(y)$ and $f_Z(z)$, respectively. If $\Theta_X, \Theta_Y$ and $\Theta_Z$ denote the circular (rotational) variables defined on the unit circle $\mathbb{S}^1$, let the PDFs be denoted as $f_{\Theta_X}(\theta_x), f_{\Theta_Y}(\theta_y)$ and $f_{\Theta_Z}(\theta_z)$, respectively. The cumulative distribution function (CDF) of the variables are represented as $F_X(X), F_Y(Y), F_Z(Z)$ for the linear variables and $F_{\Theta_X}(\theta_x), F_{\Theta_Y}(\theta_y), F_{\Theta_Z}(\theta_z)$ for the circular variables.

Based on the preliminary analysis of the data, and the histograms of the variables provided in Figures \ref{fig: c1 hist} and \ref{fig: c2 hist}, for the translational variables, the normal (N) distribution and a two component mixture of the normal (MN) distribution were considered to be most appropriate. A two component mixture of the normal distribution was considered to accommodate for the multimodality observed in the data. The expression for a finite mixture model is given as

\begin{equation}\label{eq: mixture}
    f_{X}(x) = \sum_{j=1}^{m}\omega_{j}f(x|\boldsymbol{\beta_{j}}),
\end{equation}
where $m$ is the number of mixture components, $0 < \omega_j \leq 1$ with $\sum_{j=1}^{m}\omega_{j}= 1$ representing the mixing proportions and $\boldsymbol{\beta_{j}}$ is the $j^{th}$ parameter components of the finite mixture distribution.

For the rotational variables we considered various circular distributions and found the wrapped Cauchy (wC) and a two component mixture of the wrapped Cauchy (MwC) to be most appropriate. The peakedness of the wC makes it an appealing choice for this study. The PDF of the wC\cite{mardia2009directional} is given by

\begin{equation}\label{eq: wC}
    f_\Theta(\theta) =  \frac{1}{2\pi}\frac{1 - \kappa^2}{1 + \kappa^2 - 2\kappa\cos(\theta - \mu)},
\end{equation}
where $0 \leq \kappa < 1$ and $\mu \in [-\pi,\pi)$.

To build our model we consider the defined linear and circular distributions as the marginal distributions for our framework based on the analysis of the data (see Figures \ref{fig: c1 hist} and \ref{fig: c2 hist} and Tables \ref{tab:lin stats} and \ref{tab:circ stats}). Various other distributions may also be considered depending on the complexity of a data set.

\subsection{Copulas}

The copula approach allows us to consider the marginal distributions separately from the dependence structure between the variables. Consider $F_{X_1, X_2, ..., X_d}(x_1, x_2, .., x_d)$ and  $f_{X_1, X_2, ..., X_d}(x_1, x_2, .., x_d)$ to be the joint CDF (JCDF) and joint PDF (JPDF) of the $d-$dimensional random variables $(X_1, X_2, ..., X_d)$, respectively. Then we define the copula, $C$, as
\begin{equation}\label{eq: copula func}
    F_{X_1, X_2, ..., X_d}(x_1, x_2, .., x_d) = C(u_1, u_2, ..., u_d),
\end{equation}
and thus
\begin{equation}\label{eq: copula pdf}
    f_{X_1, X_2, ..., X_d}(x_1, x_2, .., x_d) = c(u_1, u_2, ..., u_d)\prod_{j=1}^{d}f_{X_j}(x_j),
\end{equation}
 where $u_j = F_{X_j}(x_j), j= 1,2,...,d$ with $x_j \in \mathbb{R}$, $u_j \in [0,1]$ and $f_{X_j}(x_j)$ the marginal PDF of each variable.

In the linear domain, various multivariate copula functions have been defined\cite{nelsen2007introduction}. Since in directional statistics copula functions are limited to the bivariate case, we consider the use of vine copulas.

For the L-L pair copula we consider the conventional Farlie-Gumbel-Morgenstern (FGM) defined as
\begin{equation}\label{eq: FGM}
    c_{X_1, X_2}(u_1, u_2) = 1 + \alpha(1-2u_1)(1-2u_2),
\end{equation}
where $\alpha \in [-1,1]$.

For the C-L pair copula we consider the most common function proposed by Johnson et al.\cite{johnson1978some}, the JW copula, defined as
\begin{equation}\label{eq: JW copula}
    c_{\Theta, X}(u_1, u_2) = 2\pi g(\eta),
\end{equation}
where 
\begin{equation}\label{eq: eta exp}
\eta= 2\pi (u_1 - q u_2),
\end{equation}
and $\eta \in [0, 2\pi]$ is a circular random variable with $g(\cdot)$ defined as a circular PDF with $q \in \{-1,1\}$.

Jones et al.\cite{jones2015class} proposed a similar approach to Johnson et al.\cite{johnson1978some} to obtain a bivariate circular copula. The resulting copula function simplifies to a circular PDF where the argument is defined as in (\ref{eq: eta exp}).

\subsection{Proposed model}

Based on the concept of vine copulas our proposed model is built using the L-L, C-L and C-C pair copulas and their respective marginal distributions. 

\begin{figure*}[ht!]
\centering
\includegraphics[width=\textwidth]{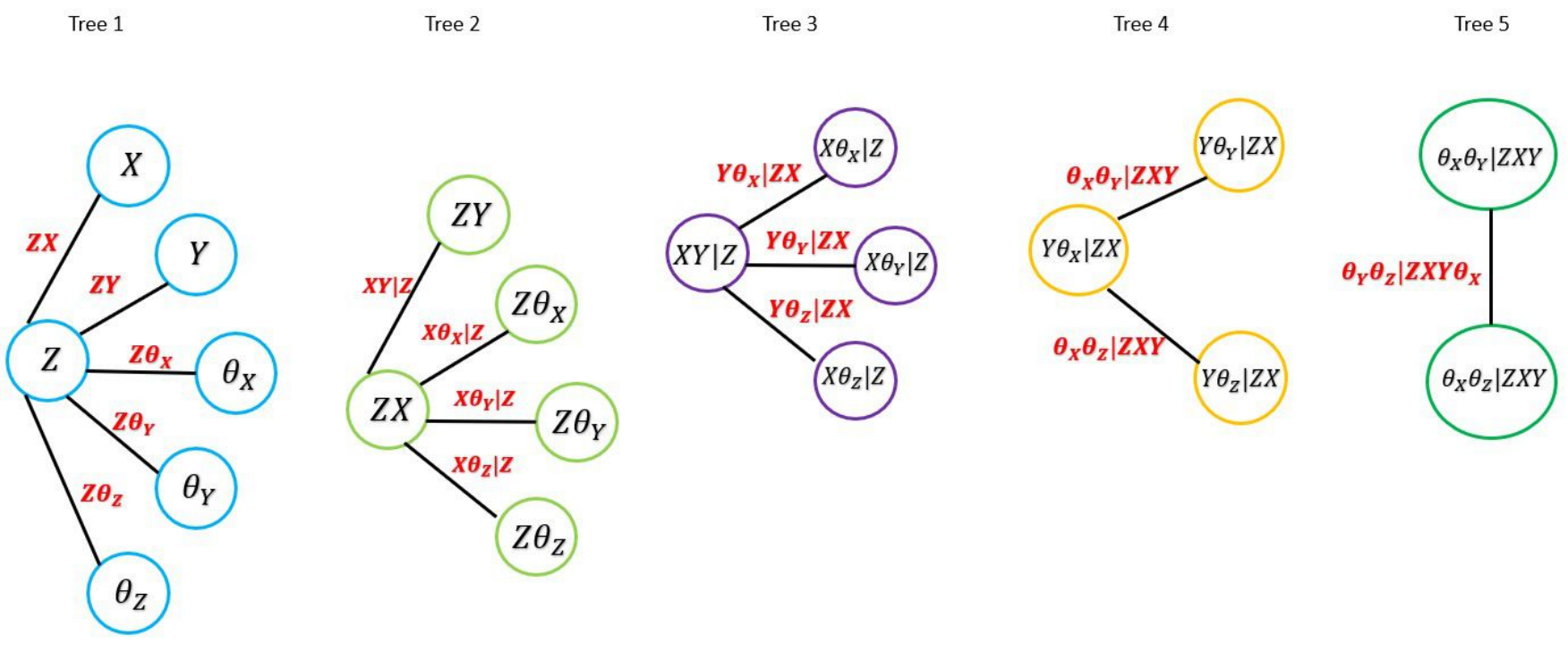}
\caption{\label{fig:tree}A schematic diagram of the canonical vine structure considered for the proposed model.}
\end{figure*}

In Figure \ref{fig:tree} the illustrated canonical vine structure considered is given. The paired variables that form the basis of the vine copula structure are denoted as $ZX, ZY, Z\Theta_X, Z\Theta_Y$ and $Z\Theta_Z$ for Tree 1 in Figure \ref{fig:tree}. The same notation follows for Trees 2 to 5 to indicate the paired variables. From the structure in Figure \ref{fig:tree} the resulting joint PDF can be extracted as follows:

\begin{align}\label{eq: jpdf}
    f(x,y,z,\theta_x,\theta_y,\theta_z) &= f(x)f(y)f(z)f(\theta_x)f(\theta_y)f(\theta_z) \nonumber \\
   & \times c_{xz}(F_X,F_Z)c_{yz}(F_Y,F_Z)c_{z\theta_x}(F_Z,F_{\theta_X})\nonumber \\
   & \times c_{z\theta_y}(F_Z,F_{\theta_Y})c_{z\theta_z}(F_Z,F_{\theta_Z}) \nonumber \\
   & \times c_{xy|z}(F_{X|Z},F_{Y|Z})c_{x\theta_x|z}(F_{X|Z},F_{\theta_X|Z})\nonumber \\
   & \times c_{x\theta_y|z}(F_{X|Z},F_{\theta_Y|Z})c_{x\theta_z|z}(F_{X|Z},F_{\theta_Z|Z}) \nonumber \\
    & \times  c_{y\theta_x|z}(F_{Y|Z},F_{\theta_X|Z})c_{y\theta_y|z}(F_{Y|Z},F_{\theta_Y|Z}) \nonumber \\
    & \times c_{y\theta_z|z}(F_{Y|Z},F_{\theta_Z|Z})c_{\theta_x\theta_y|z}(F_{\theta_X|Z},F_{\theta_Y|Z}) \nonumber \\
   & \times c_{\theta_x\theta_z|z}(F_{\theta_X|Z},F_{\theta_Z|Z})c_{\theta_y\theta_z|z}(F_{\theta_Y|Z},F_{\theta_Z|Z}).
\end{align}

It is important to note that different vine structures may lead to different results. As mentioned by Aas et al.\cite{aas2009pair}, the decomposition should be selected by determining which pair relationships are most important. Based on expert opinion, in this application we considered a structure where the variable, $Z$, is identified as most important and is linked to all the variables and is considered the conditional variable.

For the parameter estimation of (\ref{eq: jpdf}) the method of maximum likelihood estimation (MLE) is considered. Closed form expressions for the MLEs cannot be obtained due to the complex functional form. Advanced optimisation algorithms are required to numerically compute the maximum likelihood and thus the MLEs. 

\section{Application}\label{sec: application}

In this section we illustrate the validity of our proposed model by considering the fracture displacement data discussed in Section \ref{sec: motivation}.

The flexibility offered by vine copulas come at a computational cost for high dimensions. As a result, we consider a \textit{simplified} form of the joint PDF given in (\ref{eq: jpdf}) namely a truncated vine copula. For the purposes of this study we considered $K=2$ for the truncation of our vine copula. Our motivation for the choice of $K=2$ stems from the dependence structure desired, the correlations observed from the data analysis as well as the practical implications of the relationships between the variables and their interactions.

To compute our joint distribution model we first need to specify the marginal and copula functions for each variable and variable-pair, respectively. Based on preliminary analysis of the data (as illustrated in Section \ref{sec: motivation}) and models defined in Section \ref{sec: methods}, for the linear variables we consider the N distribution as well as the MN distribution. For the circular variables we consider the wC distribution as well as the MwC distribution. A combination of different marginal functions are considered for each configuration termed \textit{cases}. Table \ref{tab:app cases} provides a summary of the different cases evaluated for each configuration. We consider two cases, A and B, for each configuration ($1$ and $2$). The cases are labelled with the case option and configuration, for example, the second case of configuration 1 is denoted as case B1. The cases were chosen in consultation with domain experts and the need for a parsimonious model.

\begin{table*}[ht!]
	\small\sf\centering
	\caption{\label{tab:app cases}Combination of marginal and copula functions considered for each configuration.}
	\begin{tabular}{lllllllllllll}
		\hline
		&       & \multicolumn{11}{l}{\textbf{Function}} \\ 
		\textbf{Configuration}      & \textbf{Case} & \textbf{$X$} & \textbf{$Y$} & \textbf{$Z$} & \textbf{$\Theta_X$} & \textbf{$\Theta_Y$} & \textbf{$\Theta_Z$} & \textbf{$ZX$} & \textbf{$ZY$} & \textbf{$Z\Theta_X$} & \textbf{$Z\Theta_Y$} & \textbf{$Z\Theta_Z$} \\ 
		\hline
		\multirow{2}{*}{\textbf{1}} & \textbf{A1}    & N   & N    & N    & wC    & wC     & wC   & FGM   & FGM    & wC (JW)  & wC (JW) & wC (JW) \\ \cline{2-13} 
		& \textbf{B1}  & N  & N  & N   & MwC  & wC   & MwC   &  FGM   &  FGM  & wC (JW)  & wC (JW) & wC (JW) \\ \hline
		\multirow{2}{*}{\textbf{2}} & \textbf{A2}    & N  & N   & N   & wC    & wC  & wC    & FGM  & FGM & wC (JW)  & wC (JW) & wC (JW)  \\ \cline{2-13} 
		& \textbf{B2}  & N   & MN   & N   & wC   & wC   & MwC  & FGM  & FGM   & wC (JW)   & wC (JW)  & wC (JW) \\
		\hline
	\end{tabular}
\end{table*}

For the finite mixture models, the expectation-maximisation (EM) algorithm may be utilised. Due to the high dimensional space of the parameter set, advanced optimisation algorithms such as the particle swarm optimisation\cite{wang2018particle} (PSO) and genetic algorithm\cite{chatterjee1996genetic} (GA) were used to efficiently estimate the parameters of the joint model. 

In Table \ref{tab:perofmance measures} the performance measures, for the different cases specified in Table \ref{tab:app cases}, are provided. For the performance evaluation two goodness-of-fit metrics are applied to evaluate the models. The Akaike information criterion (AIC) estimates the relative amount of information lost by a given model. The Bayesian information criterion (BIC) is widely used for model selection and is similar to the AIC, however, the BIC is more strict in its penalisation of model complexity. These two metrics are defined as follows,

\begin{equation*}
    AIC =  2p - 2\ln(\hat{L}),
\end{equation*}
and 
\begin{equation*}
    BIC  =  p\ln(n) - 2\ln(\hat{L}),
\end{equation*}

\noindent where $p$ is the number of estimated parameters in the model, $n$ the total number of data points, and $\hat{L}$ the maximum value of the likelihood function for a specific~model. Based on the performance measures in Table \ref{tab:perofmance measures}, we can conclude that case B1 and case B2 are the best models for configuration 1 and configuration 2, respectively.

\begin{table*}[ht!]
	\small\sf\centering
	\caption{\label{tab:perofmance measures}The number of parameters ($p$), Maximised log-likelihood (MLL), Akaike information criterion (AIC) and Bayesian information criterion (BIC) values for the different combinations of marginal and copula functions.}
	\begin{tabular}{llllll}
		\hline
		\textbf{Configuration}      & \textbf{Case} & \textbf{$p$} & \textbf{MLL} & \textbf{AIC} & \textbf{BIC} \\ 
		\hline
		\multirow{2}{*}{\textbf{1}} & \textbf{A1}    &  17  &    169.4548   &   -304.9096     &   -264.4152   \\ \cline{2-6} 
		& \textbf{B1}  &  23  &  170.6057  &  -295.2114   &  -240.4248 \\ \hline
		\multirow{2}{*}{\textbf{2}} & \textbf{A2}    &  17  &  37.7123     & -41.4246  & -0.9302   \\ \cline{2-6} 
		& \textbf{B2}    &  23  & 103.2792    & -160.5584    & -105.7718   \\ 
		\hline
	\end{tabular}
\end{table*}

\begin{remark}
It is important to note that the results of the proposed model cannot be compared with existing methods (defined in the linear domain) as the models are defined on different manifolds. Due to the nature of the rotational variables, implementing an approach that incorporates directional statics is crucial for accurate modelling.  
\end{remark}

To illustrate the validity of the proposed model in comparison to the conventional use of an independent model, we consider a likelihood ratio test (LRT). We consider the model under the null hypothesis to be the independent model and case B1 and case B2 (models of best fit) to be the model under the alternative hypothesis, respectively. For the independent model, we consider the same distributions as specified for case B1 and case B2 for the six variables ($X, Y, Z, \Theta_X, \Theta_Y$ and $\Theta_Z$), respectively. For configuration 1, independent model vs. case B1, we obtain a LRT value of $80.4512$ and reject the null hypothesis at a 5\% significance level (p-value = $6.66e^{-16}$). For configuration 2, independent model vs. case B2, we obtain a LRT value of $22.1448$ and reject the null hypothesis at a 5\% significance level (p-value = $0.0005$). Thus, we can conclude that the joint dependent model is a more suitable fit for the data. 

\section{Conclusion}\label{sec: conclusion}

In this paper we propose a modelling framework applicable to the 6D joint distribution. This model accounts for the cyclicity of the rotational variables by means of directional statistics as well as accounts for a dependence structure between the variables. The framework is constructed based on vine copulas. The pair copula decomposition concept of vine copulas represents the dependence structure as a combination of circular-linear, circular-circular and linear-linear pairs modelled by their respective copulas. This allows us to assess the dependencies in the joint distribution. An advantage of using vine copulas is the flexibility to build multivariate distributions via bivariate copulas that model the dependence between pairs of random variables. For efficient estimation a truncation of the vine copula was considered. The analysis of this data motivates the need for a dependence structure to be accounted for when modelling this type of data. From the results of the real data application, the advantage of applying the joint dependence model is observed. Based on the likelihood ratio test we can conclude that the joint dependent model is a better choice for modelling the fracture displacements and will thus be more informative for evaluations of these devices and the design thereof. The proposed modelling framework can be adjusted for other practical cases depending on the desired dependency structure required and the relationship between the variables. The primary goal of an external fixator is injury rehabilitation. Fracture healing is inevitably influenced by the complex interplay of biology and biomechanics - i.e. inter-fragmentary motion and biomechanics. The construct of an external fixator is determined by the configuration of the hardware. The construct’s configuration may lead to different inter-fragmentary motions and it is therefore important to accurately model and understand the motions in play to obtain the most appropriate construct for optimal healing. The modelling framework proposed in this paper provides a more accurate 
view for fracture displacements thus leading to improved evaluations and design of these devices, thus, aligning with the United Nation's Sustainable Development Goal (SDG) 3 to promote good health and well-being. 

\vspace{0.8cm}

\section*{Funding}
This work was based upon research supported in part by the National Graduate Academy for Mathematical and Statistical Sciences (NGA) of South Africa; National Research Foundation (NRF) of South Africa, Reference: SRUG2204203965, grant no. 120839 and Reference: RA171022270376, grant no. 119109;  DSI-NRF Centre 386 of Excellence in Mathematical and Statistical Sciences (CoE-MaSS), South Africa and STATOMET at the Department of Statistics at the University of Pretoria. The research of M. Arashi is based upon research funded by Iran National Science Foundation, grant NO. 4015320.

\bibliographystyle{plain}
\bibliography{biomechincal_ref}

\end{document}